\documentclass[12pt,a4paper]{article}
\usepackage[utf8]{inputenc}
\usepackage{color}
\usepackage{parskip}
\usepackage{enumitem}
\usepackage{amsmath,amscd}
\usepackage{stmaryrd}
\usepackage{mathrsfs}
\usepackage{amsfonts}
\usepackage{amssymb,scalerel,stackengine}
\usepackage[most]{tcolorbox}
\newtcbox{\othermathbox}[1][]{nobeforeafter, math upper, tcbox raise base, enhanced, rounded corners,colback=black!3, colframe=black}
\usepackage{empheq}
\usepackage[margin=1in]{geometry}
\usepackage{graphicx}
\usepackage[leftcaption]{sidecap}
\usepackage[font=small]{caption}
\usepackage{subcaption}
\usepackage{here}
\usepackage[colorlinks=true,allcolors=blue]{hyperref}
\usepackage{cite}
\usepackage[british]{babel}
\usepackage{hhline}
\usepackage{multirow}
\usepackage{changepage}
\usepackage{comment}

\renewcommand{\thesection}{\arabic{section}}

\numberwithin{equation}{section}

\newcommand\beq{\begin{align}}
\newcommand\ee{\end{align}}

\newcommand{\Comment}[1]{}
\newcommand\dd{\mathrm{d}}

\newcommand*{\vrectangleA}{{\ooalign{\lower.3ex\hbox{$\sqcup$}\cr\raise.4ex\hbox{$\sqcap$}}}}

\newcommand{\overbar}[1]{\mkern 1.5mu\overline{\mkern-1.5mu#1\mkern-1.5mu}\mkern 1.5mu}

\def\ie{{\emph{i.e. }}}
\def\eg{{\emph{e.g. }}}

\def\hi{{\hat i}}
\def\hj{{\hat j}}
\def\hk{{\hat k}}

\def\hmu{{\hat \mu}}
\def\hnu{{\hat \nu}}

\def\ha{{\hat A}}
\def\hb{{\hat B}}

\def\hmu{{\hat \mu}}
\def\hnu{{\hat \nu}}
\def\hr{{\hat r}}
\def\mn{{\mu\nu}}

\def\l{{\lambda}}
\def\L{\Lambda}
\def\de{\delta}

\def\pd{\partial}

\def\cd{\nabla}
\def\eps{\epsilon}

\def\cM{\mathcal{M}}

\def\cP{\mathcal{P}}

\def\cO{\mathcal{O}}

\def\th{\theta}

\def\tl{\widetilde}

\newcommand\bolde{{\boldsymbol{e}}}
\newcommand\be[1]{{\boldsymbol{e}_{\hat{#1}}}}

\newcommand\e[2]{{E_{\hat{#1}}^{\;\;#2}}}

\newcommand\barE[2]{{E_{\hat{#1}}^{\;\;#2}}}

\def\deq{{\;\dot{=}\;}}

\title{\bf Memory effects from holonomies}
\author{Ali Seraj and Turmoli Neogi\\[1em]
	{\small Centre for Gravitational Waves, Université Libre de Bruxelles,}\\[-.2em]
	{\small International Solvay Institutes, CP 231, B-1050 Brussels, Belgium.}}

\date{}

\begin{document}

\maketitle

\begin{center}
	\begin{minipage}{.85\textwidth}
		{\bf Abstract.} We provide a uniform treatment of electromagnetic and gravitational memory effects, based on the gravito-electromagnetic formulation of GR and a generalization of the geodesic deviation equation. This allows us to find novel results: in gauge theory, we derive relativistic corrections to the well-known kick memory observable, and a general expression for the displacement memory observable, typically overlooked in the literature. In GR, we find relativistic corrections to displacement and kick memory observables. In both theories, we find novel radial memory effects. 
		
		Next, we  show that electromagnetic and gravitational memory observables can be formulated in terms of certain holonomies on a holographic screen in asymptotically flat spacetimes. 
		In gauge theory, the displacement and kick memory effects form a Hamiltonian vector field which is canonically generated by a Wilson loop. In the first order formulation of GR, we show that the holonomy naturally splits into translational and Lorentz parts. While the former encodes the leading and subleading displacement and kick memory observables, the latter reproduces the gyroscopic memory effect. 
			\end{minipage}
\end{center}
\tableofcontents
\section{Introduction and summary of results}
Gravitational memory refers to the \textit{plasticity} of spacetime subjected to gravitational waves (GWs). This leads to permanent effects in the configuration of experimental setups, such as  a change in the physical distance between free test masses which are initially at rest \cite{Zeldovich:1974gvh,Braginsky:1985vlg}, or a net rotation in the orientation of a gyroscope\cite{Seraj:2021rxd,Seraj:2022qyt}. 
The former is the well known `displacement memory' whose linear form was discovered in the 70's . A more rigorous analysis of this effect in the 90's by Christodoulou \cite{Christodoulou:1991cr} and Blanchet and Damour \cite{Blanchet:1992br} showed that there is a nonlinear contribution to the memory, which can be understood as the linear memory caused by emitted gravitons \cite{Thorne:1992sdb}, \ie   the ability of gravitons to gravitate\cite{Favata:2010zu}. While the linear effect is dominant in the memory caused by scattering processes, the nonlinear Christodoulou effect has the major contribution in bound systems, such as binary coalescence \cite{Wiseman:1991ss}. Thanks to the BMS flux-balance equations\cite{Compere:2019gft}, memory effects can be implemented in numerical waveforms, implying that the effect can be enhanced in the merger phase \cite{Mitman:2020bjf,Mitman:2020pbt,Mitman:2021xkq}. However, the low frequency nature of the memory, makes it hard to detect in GW experiments. Prospects for detection of GW memory is discussed in \cite{Favata:2010zu,Lasky:2016knh,Hubner:2019sly,Boersma:2020gxx,Hubner:2021amk,Grant:2022bla}.  

In recent years, gravitational memory effect has attracted new interest, due to its fundamental relationship with the symmetry structure of  asymptotically flat spacetimes \cite{Strominger:2014pwa,Pasterski:2015tva,Strominger:2021lvk,Strominger:2021mtt,Freidel:2021dfs,Geiller:2022vto}. The leading displacement memory is proportional to the permanent change in the \textit{shear}, \ie the transverse-traceless components of the metric, which can be interpreted as a `vacuum transition' under the action of BMS supertranslations \cite{Strominger:2014pwa,Himwich:2019qmj,Godazgar:2022pbx}. Recently, new types of permanent GW effects have been discovered that are not simply related to the net change in the shear, but are rather sourced by certain time integrals of the waveform. This includes the spin and center of mass memory effects \cite{Pasterski:2015tva,Nichols:2017rqr,Nichols:2018qac}, and the gyroscopic memory effect \cite{Seraj:2021rxd,Seraj:2022qyt}. Generalized memory effects can be described uniformly in terms of various integer modes in the Mellin transform of the Bondi news \cite{Freidel:2021ytz,Compere:2022zdz} and can be observed through the ``curve deviation'', a generalization of the geodesic deviation \cite{Grant:2021hga}.  While all these effects are subdominant with respect to the leading displacement memory and thereby hard to detect in the near future, they are still very interesting due to their relation to the complete symmetry structure of gravity, which includes the loop algebra of $w_{1+\infty}$ \cite{Strominger:2021mtt,Freidel:2021ytz, Compere:2022zdz}. Finally, it should be mentioned that there are novel memory effects in modified theories of gravity \cite{Tahura:2020vsa,Tahura:2021hbk,Hou:2020tnd} with interesting interpretation in terms of \textit{dual} charges and symmetries \cite{Seraj:2021qja}.

Memory effects also show up  in gauge theories. It is well known that  the passage of electromagnetic radiation leads to a net shift in the transverse components of the gauge field, which leads to a ``kick''(change of velocity) in a charged particle \cite{Bieri:2013ada,Susskind:2015hpa, Pasterski:2015zua}. We revisit this problem and show that there is also a subleading displacement effect in gauge theory, which  generalizes previous result of \cite{Mao:2017axa}. In this context, similarities and differences between gravity theory and gauge theory become much more transparent.

Memory is a nonlocal effect both in space and time. Therefore, one might expect that nonlocal quantities like holonomies and Wilson loops be good candidates to quantify memory effects. This is supported by results in \cite{Pate:2017vwa,Ball:2018prg,Choi:2019fuq,Choi:2019sjs}, where it is shown that final states in a scattering process are dressed by Wilson lines anchored to the celestial sphere at infinity. A more concrete result on the relation between memory and holonomy in GR was achieved in \cite{Flanagan:2014kfa,Flanagan:2018yzh} where a `generalized'  holonomy was introduced through a modification of the parallel transport equation. It was shown that this holonomy contains displacement and velocity  memory effects.

\paragraph{Summary of results and outline.}
In this paper, we provide a uniform treatment of electromagnetic and gravitational memory effects, using a generalized Lorentz force equation given in \eqref{generalized Lorentz force}. From this, we derive the general result \eqref{memories general} for the displacement and kick memory effects, which applies both to electromagnetism (EM) and general relativity (GR). One can easily specialize this result to EM or GR by identifying the transverse (gravito)electric field, leading respectively to \eqref{memories EM} and \eqref{gravitational memories}. We will discuss several improvements of our results with respect to the existing literature. 

Next, we show that there is a correspondence between memory and holonomy. We provide an explicit expression to extract both leading and subleading displacement and kick memory effects from the holonomy. This applies both to EM and GR. In case of EM, we show that the displacement and kick memory effects form a Hamiltonian vector field in the phase space of the test particle, which is canonically generated by the Wilson loop, see Eq.\eqref{memory-holonomy-canonical}. Similar result is obtained in Eqs.\eqref{holonomy displacement memory}, \eqref{Memory-holonomy-GR} for the gravitational case, but we have not been able to identify a canonical structure in this case.

One novelty of our approach with respect to that of \cite{Flanagan:2014kfa,Flanagan:2018yzh} is that we study the gravitational holonomy in terms of tetrad variables. In this setup, the holonomy is an element of the Poincar\'e group, which naturally splits into into translation and Lorentz contributions. We develop this in section \ref{sec: GR holonomy formal}, and show that the translational holonomy obeys the affine transport equation proposed in  \cite{Flanagan:2014kfa,Flanagan:2018yzh}. Then, we will show in section \ref{sec: translational holonomy} that the translational holonomy reveals displacement and kick memories, while the rotational holonomy is shown to reproduce the gyroscopic memory effect in section \ref{sec:rotational holonomy}. We conclude in section \ref{sec:discussion}. Throughout the paper, we will use geometrized units in which the speed of light $c=1$.

\section{EM Memory Effects}\label{sec:Maxwell}
\subsection{Asymptotic analysis}
Consider a test particle with mass $m$ and charge $q$, in Minkowski space, subject to electromagnetic radiation produced by some source $J^\mu$ at a large distance $r$. We assume that the source is of compact spatial support, centered around the origin $(r=0)$ of the retarded coordinate system $(u,r, \th^A)$ in which the Minkowski metric reads
\begin{align}\label{metric}
	ds^2&=-du^2-2dudr +r^2 \gamma_{AB}d\th^Ad\th^B\,.
\end{align}
In these coordinates, Maxwell equations $	\pd_\mu F^\mn=J^\nu$ imply that the gauge field behaves at large distance as \cite{Madler:2016xju}
\begin{align}\label{falloff gauge field}
	A =\frac{\overbar{A}_u}{r} du + \frac{\overbar{A}_r}{r^2} dr + \overbar{A}_A d\th^A+\text{subleading}\, . 
\end{align}
Note that coordinate basis vectors $\pd_A$ are not normalized, since $\pd_A\cdot\pd_B=r^2 \gamma_{AB}$. 
From the viewpoint of an experimenter, it is more natural to use instead an orthonormal tetrad, \ie  a set of four basis vectors $\boldsymbol{e}_\hmu=e_{\hmu}{}^{\mu}\,\pd_\mu$ such that $\boldsymbol{e}_\hmu\cdot\boldsymbol{e}_\hnu=\eta_{\hmu\hnu}$ where $\eta_{\hmu\hnu}=\text{diag}({-1,1,1,1})$. The dual basis one forms are given by $\boldsymbol{e}^\hmu={e}^\hmu{}_\mu dx^\mu$. 
In this frame, components of any tensor is given by contraction with suitable number of the tetrad $e_{\hmu}{}^{\mu}$ or its inverse $e^{\hmu}{}_{\mu}$. For example, $A_\hmu=A_\mu e_\hmu{}^\mu$ or $F^{\hmu\hnu}=e^\hmu{}_{\mu}e^\hnu{}_{\nu}F^{\mn}$.   A suitable frame in  Minkowski spacetime is 
\begin{align}\label{frame Minkowski}
	\be{0}=\pd_u,\quad \be{r}=\pd_r-\pd_u,\quad \be{A}=\frac{1}{r}\e{A}{A}\,\pd_A\,,\\
	\bolde^{\hat 0}=du+dr,\quad \bolde^{\hat r}=dr,\quad \bolde^{\hat A}=r E^{\hat A}{}_{A} \,d\th^A\,,
\end{align}
where $\be{0}$ is adapted to observers at rest, $\be{r}$ is along the radial direction specified by outgoing rays, and $\e{A}{A}$ form a dyad on the round sphere $\gamma_{AB}\e{A}{A}\e{B}{B}=\de_{\hat A\hat B}$. In this basis, we have 
\begin{align}\label{gauge field expansion}
		A =\frac{1}{r}\left({\overbar{A}_{\hat 0} \bolde^{\hat 0} + \overbar{A}_{\hat r} \bolde^{\hat r} + {\overbar A}_{\hat A}}\bolde^{\hat A}\right)+\cO(1/r^2)\, ,
\end{align}
where the leading components relate to those in the coordinate basis as 
\begin{align}
	\overbar{A}_{\hat 0}=\overbar{A}_u=-\overbar{A}_{\hat r}, \qquad \overbar{A}_{\hat A}=\e{A}{A}\overbar A_A \,.
\end{align}
The last equation suggests that $\overbar A_A (u,x^B)$ should be viewed as a tensor living on the unit sphere and therefore it is contracted with the dyad on the sphere $\e{A}{A}$. This will be frequently used in the paper, following the Bondi framework in which one performs an asymptotic expansion in $1/r$ to reduce dynamical fields to covariant tensors living on the unit `celestial' sphere. 
Electric and magnetic fields are defined in the local frame as: 
\begin{align}\label{E, B def}
E^\hi=F^{\hat{0}\hi}\,,\qquad B^{\hi}=\frac{1}{2}\eps^{\hi\hj\hk}F_{\hj\hk}\,,
\end{align}
 where $\hi,\hj,\hk$ are Cartesian spatial indices $(\hr,\ha)$. Simple manipulation yields 
\begin{subequations}\label{F asymptotic}
\begin{align}	
	E_{\ha}&=\frac{\overbar{E}_{\ha}}{r}+\cO(1/r^2),& B_{\ha}&=-\eps_{\ha}{}^{\hb}{{E}}_{\hb}+\cO(1/r^2)\,,\\
	E_{\hr}&=\cO(1/r^2)&	B_{\hr}&=\cO(1/r^2)\,,
\end{align}
\end{subequations}
where 
\begin{align}\label{Electric field EM}
{\overbar{E}}_{\ha}={\dot{\overbar{A}}}_{\ha}\,,\qquad\qquad\text{in EM}.
\end{align}
We will show in section \ref{sec: translational holonomy} that Eq. \eqref{F asymptotic} also holds in GR for suitable definitions of gravito-electric and gravito-magnetic fields. Therefore, the electric and magnetic fields are asymptotically transverse and orthogonal to each other.

\subsection{Generalized Lorentz force law}
In this section, we study persistent effects of  EM or gravitational waves on test bodies. In Maxwell theory, the effect of EM fields on a test charge is given by the Lorentz force law. In GR, on the other hand, displacement memory effects are typically studied using the geodesic deviation equation. In this work, however, we will work with a generalized force law, which  allows treating EM and GR effects in a unified manner. Moreover, in the gravitational case, it matches the so-called \textit{Bazanski} equation, which contains higher order relativistic corrections to the geodesic deviation equation. 

Consider an inertial observer carrying an orthonormal tetrad $\be{\mu}$ , parallel transported along the observer's worldline. The tetrad can be used to locally define a Fermi normal coordinate system $(T,X^\hi)$ such that $\be{0}=\pd_T, \be{i}=\pd_{X^\hi}$.\footnote{Since we are performing an asymptotic analysis, we need these properties, such as the parallel transport of the tetrad to hold only at leading order in $\cO(1/r)$.} In such local coordinate system, the effect of EM (gravitational) field on a test charge (mass) is given by
	\begin{empheq}[box=\othermathbox]{align}\label{generalized Lorentz force}
	\frac{d{ V}^{\hi}}{dT} =\beta_E E^\hi+\beta_B (V\times B)^\hi +F^\hi_{\text{ext}}\,,
\end{empheq}
where $V^\hi=\frac{dX^\hi}{dT}$ is the coordinate velocity of the test body. This equation resembles the non-relativistic form of Lorentz force equation, except that we allow for arbitrary coupling $\beta_{E},\beta_{M}$ to (gravito) electric and (gravito) magnetic fields, and an ``extra'' force (per unit mass) $F^\hi_{\text{ext}}$ in order to unify the analysis of EM and GR. The extra force is subleading relativistic correction to the first two effects, as will be detailed later. Moreover, its specific form depends on the theory. $F^\hi_{\text{ext}}$ also accommodates possible external forces in the problem, but we will not consider this possibility in this work.

\paragraph{Setup of the experiment.} Consider a test body with initial position and velocity $(X^\hi_0,V^\hi_0)$ in the local frame. 
We take the test body to be far from the source of radiation and thus we ignore $\cO(1/r^2)$ effects, unless otherwise stated (\eg in gyroscopic memory effect). At leading order in $1/r$ expansion, the observer's tetrad can be chosen to coincide with \eqref{frame Minkowski}, and as a result $T=u$ up to a shift in the origin of time, which we set to zero. Using the asymptotic relation \eqref{F asymptotic} between electric and magnetic fields (which holds both in EM and GR) in \eqref{generalized Lorentz force} and keeping only $1/r$ effects, we find
\begin{align}\label{Forces}
 {\dot V}^{\hr}&=\frac{1}{r} \left(\beta_B\,{\overbar{E}}_{\ha}V^\ha+f^\hr\right)\,,\qquad	{\dot V}^\ha=\frac{1}{r}\left((\beta_E-V^\hr \beta_B)\,{\overbar{E}}^\ha +f^\ha\right)\,,
\end{align}
where ${F}^\hi_{\text{ext}}=\frac{1}{r}f^\hi(u,\th^A)+\cO(1/r^2)$.
We assume that the radiation is confined in a finite time interval, \ie ${\overbar{E}}_{\ha}=0=f^\hi$ outside the (possibly long) time interval $(u_0,u_f)$, which guarantees the convergence of time integrals that will appear.\footnote{While this is a reasonable assumption in EM, it is merely an approximation in GR, due to nonlinear tail effects, leading to a power-law decay in the radiation. We will not consider this issue in this paper.} Eq.\eqref{Forces} implies that $V^\hi=V^\hi_0+\cO(1/r)$ and thus, at leading order in $1/r$, we can replace $V^\hi$ by $V_0^\hi$ on the right hand side. The position of the test body is thus given by 
\begin{align}\label{X formal}
	X^\hi(u)&=X_0^\hi+\int_{u_0}^u dv\int_{u_0}^{v} dw\dot{V}^\hi(w)\,.
\end{align}
Since the acceleration $\dot{V}^\hi$ vanishes  before the arrival and after the passage of the burst of radiation, we can write
\begin{align}
	X^\hi(u)&=X_0^\hi+{V_0}^{\hi}(u-u_0)\,,& u&\leq u_0\,,\\
	X^\hi(u)&=X_0^\hi+\Delta X^\hi+({V_0}^{\hi}+\Delta V^\hi)(u-u_0)\,,& u&\geq u_f\,.
\end{align}
The latter equation encodes the kick $\Delta{V}^{\hi}$ and displacement $\Delta{X}^{\hi}$ memory observables.\footnote{
What we call the displacement memory observable $\Delta X^\hi$ refers to radiation effects on a probe system of test particles. In the literature, sometimes ``memory effects'' (leading or subleading) refer to certain contributions in the  waveform. We will see explicitly in \eqref{memories EM}, \eqref{gravitational memories} that the displacement momory observable $\Delta X^\hi$ provides a pairing between parameters of the probe system $(X_0,V_0)$ and memory effects in the waveform, which provides an explicit relationship between the two notions. A non-geodesic curve is specified by additional acceleration moments which pair with more subleading data in the waveform through the \textit{curve deviation} observable \cite{Flanagan:2018yzh,Grant:2021hga}.}  Defining the Mellin transform of a function $f(u)$ as $\cM_s(f)\equiv\int_{u_0}^{\infty}(u-u_0)^{s-1} f $, the result can be compactly represented as
\begin{subequations}\label{memories general}
	\begin{empheq}[box=\othermathbox]{align}
		\Delta{V}_{\ha}&\deq\frac{1}{r}\,\cM_1(\beta{\overbar{E}}_\ha+f_\ha)\,,
		& \Delta{X}_{\ha}&\deq-\frac{1}{r}\,\cM_2(\beta{{\overbar{E}}}_\ha+f_\ha)\\
		\Delta{V}_{\hr}&\deq\frac{1}{r} V_0^\ha\,\cM_1(\beta_B {\overbar{E}}_\ha+f_\ha)\,,
		&\Delta{X}_{\hr}&\deq-\frac{1}{r}\, V_0^\ha \cM_2(\beta_B{{\overbar{E}}}_\ha+f_\ha)
	\end{empheq}
\end{subequations}
where $\beta\equiv \beta_E-\beta_B V_0^\hr $, and we have used several integration by parts to bring the result into this form. Therefore, we observe that the kick and displacement memory effects are respectively sourced by integer $s=1,2$ modes in the Mellin transform of radiation fields. 

\subsection{EM memory effects} 
In the case of EM, the Lorentz force equation is given by  $\frac{du^\mu}{d\tau}=\frac{q}{m}F^{\mn}u_\nu$, where $u^\mu$ is the four velocity and $\tau$ is the proper time on the particle's worldline. Expanding this in terms of the velocity with respect to a coordinate system $(T,X^\hi)$, we find 
\begin{align}
	\frac{d{ V}^{\hi}}{dT} =\frac{q}{m\gamma}\left[F^{\hat{0}\hj}(\de^\hi{}_{\hj}-V^\hi V_\hj)+F^{\hi\hj}V_{\hj}\right]
\end{align}
where $\gamma=(1-|V|^2)^{-1/2}$ is the Lorentz factor. Having \eqref{E, B def} in mind, we recover \eqref{generalized Lorentz force} with 
\begin{align}
\beta_B=\beta_E=\frac{q}{m\gamma}\,,\qquad F_{\text{ext}}^\hi=	-\beta_{E} V^\hi V^\hj E_\hj \,.
\end{align}
Therefore, \eqref{memories general} and the asymptotic form \eqref{Electric field EM} imply that up to $\cO(r^{-2})$ corrections, 
\begin{subequations}\label{memories EM}
	\begin{align}
		\Delta{V}^{\ha}&=\frac{q}{m\gamma_0 r}\left(\de^{\hat A\hat B}(1-V_0^\hr )-V_0^\ha V_0^\hb\right)\,\Delta{\overbar{A}}_\hb\,,\\
		\Delta{X}^{\ha}&=\frac{q}{m\gamma_0 r}\left(\de^{\hat A\hat B}(1-V_0^\hr )-V_0^\ha V_0^\hb\right)\,\int_{u_0}^{u_f} du\big({\overbar{A}}_\ha(u)-{\overbar{A}}_\hb(u_f)\big)\,,\label{displacement EM}\\
		\Delta{V}^{\hr}&=\frac{q}{m\gamma_0 r}\, (1-V_0^\hr)\, V_0^\ha\,\Delta{\overbar{A}}_\ha\,,\label{radial kick}\\
		\Delta{X}^{\hr}&=\frac{q}{m\gamma_0 r}\,(1-V_0^\hr)\, V_0^\ha\int_{u_0}^{u_f} du\big({\overbar{A}}_\ha(u)-{\overbar{A}}_\ha(u_f)\big)\,.	\label{radial displacement}
	\end{align}
\end{subequations}
where $\Delta{\overbar{A}}_\ha \equiv\lim_{u\to\infty}{\overbar{A}}_\ha(u)-{\overbar{A}}_\ha(-u)={\overbar{A}}_\ha(u_f)-{\overbar{A}}_\ha(u_0)$ and $\gamma_0\equiv\gamma(V_0)$. Let us compare this result with the literature. The transverse component of the kick memory matches with the result of \cite{Bieri:2013hqa}, in the non-relativistic limit $|V_0|\to 0$. On the other hand, the displacement effect was found in \cite{Mao:2017axa}, in the special case, where the gauge field has odd parity.  Radial memory effects and relativistic corrections to all these quantities are new in our result, as initial velocity is not considered in the previous literature on EM memory effects (up to our knowledge). Note that radial memory effects are suppressed with respect to transverse effects by a factor $|V_0|/c$.

\section{Gravitational memory effects}\label{sec:GR memory effects}
One can consistently incorporate gravitational effects, including gravitational radiation, in the Minkowski metric \eqref{metric}. This is conveniently done in Bondi gauge \cite{Bondi:1962px,Sachs:1962wk}
\begin{align}
\label{bog}
g_{rr}=g_{rA}=0,
\qquad
\pd_r\det\left(r^{-2}g_{AB}\right)=0\,,
\end{align}
in which any metric can be written as \cite{Barnich:2009se,Barnich:2010eb}
\begin{align}
	\label{e2}
	ds^2
	=
	-e^{2\beta}(F d u^{2}+2 du\,dr)
	+r^2h_{AB}\big(d\th^{A}-\frac{\,U^{A}}{r^2} du\big)\big(d\th^{B}-\frac{\,U^{B}}{r^2}du\big)\,,
\end{align}
where $F,\beta, U^A, h_{AB}$ are functions of all coordinates. To understand the structure of radiative spacetimes, one solves Einstein equations far from the source, treating $1/r$ as a small parameter and imposing as a boundary condition that the metric tends to the Minkowski metric in the limit $r\to \infty$, \ie $\lim_{r\to\infty}h_{AB}=\gamma_{AB}$, where $\gamma_{AB}$ is the round metric on the unit `celestial sphere'. At leading order, one finds
\begin{align}
	\dd s^2\label{Bondi metric}
	=&-\big(1-\tfrac{2m}{r}\big)\dd u^2-2\dd u\,\dd r+\big(r^2\gamma_{AB}+rC_{AB}\big)\dd\th^A\,\dd\th^B
	+D^B C_{AB}\dd u\,\dd\th^A\,.
\end{align}
The symmetric trace-free tensor $C_{AB}(u,\th^A)$, called the Bondi shear, is the shear of the congruence of outgoing null rays and encodes the gravitational waveform received at retarded time $u$ and angles $\th^A$ on the celestial sphere. On the other hand, $m(u,\th^A)$ is the Bondi mass aspect whose flux is fixed by the radiation $\dot m=\frac{1}{4} D_{A} D_{B} \dot{C}^{A B}-\frac{1}{8} \dot{C}_{A B} \dot{C}^{A B}$. 
Note that $\gamma_{AB}$ and its inverse $\gamma^{AB}$ are used to lower or raise indices of tensors living on the celestial sphere and $D_A$ is the associated covariant derivative, $D_C\gamma_{AB}=0$.
\subsection{Displacement and kick memory}\label{sec:GR memories}
The geodesic deviation equation, typically used to study displacement memory effect\cite{Christodoulou:1991cr}, describes the relative separation of a nearby freely falling test mass with respect to a reference freely falling observer. In a Fermi normal coordinate system adapted to the observer described in the previous section, the geodesic deviation equation reads
\begin{align}
\frac{d^2{X}^{\hi}}{dT^2} +R^{\hi}{}_{\hat{0}\hj\hat{0}} \, X^{\hj}=O(X,\dot{X})^2\,,
\end{align}
where $O(X,\dot{X})^n$ refers to corrections containing at least $n$ factors of $X^\hi$ and/or $\dot{X}^\hi$. However, a more careful treatment reveals the \textit{Bazanski} equation (see (A.12) of \cite{Vines:2014oba} and (4.5) of \cite{Flanagan:2018yzh} for a modern derivation and references therein for original works):
\begin{align}\label{Bazanski}
\frac{d^2{X}^{\hi}}{dT^2}  +R^{\hi}{}_{\hat{0}\hj\hat{0}} \, X^{\hj}+2R^{\hi}{}_{\hk\hj\hat{0}}X^\hj \dot{X}^\hk+\cd_{(\hat{0}}R_{\hk)\hi\hat{0}\hj}X^\hj X^\hk =O(X,\dot{X})^3\,.
\end{align}
We assume in this section that $X^\hi$ is small with respect to other length scales in the problem, including the wavelength of the radiation and the distance to the source, and that the velocity $V^\hi$ is small with respect to that of light, so that we ignore subleading $O(X,\dot{X})^3$ corrections. The Bazanski equation thus coincides with the generalized Lorentz force law \eqref{generalized Lorentz force}, with
\begin{subequations}
\begin{align}\label{GEM fields}
	E_\hi&=-R_{\hat{0} \hi \hat{0} \hj} X^\hj\,,&B_\hi&=\frac{1}{2} \epsilon_{\hi \hj \hk} R^{\hj \hk}{}_{\hat{0} \hat{l}} X^{\hat l}\,,&	F_{\text{ext}}^\hi&=-\cd_{(\hat{0}}R_{\hk)\hi\hat{0}\hj}X^\hj X^\hk\,,\\
	\beta_E&=\frac{q_E}{m}=1,&\beta_B&=\frac{q_B}{m}=2\,.&
\end{align}
\end{subequations}
Here $E^\hi,B^\hi$ are the gravito-electric and gravito-magnetic fields. We observe the important difference between EM and GR: while $\beta_{B}/\beta_{E}=1$ in EM, $\beta_{B}/\beta_{E}=2$ in GR\cite{Mashhoon:2003ax}. 

In an asymptotically flat spacetime, Eq.~\eqref{frame Minkowski} still defines an asymptotically orthonormal basis, based on which one can construct a local coordinate system $(T,X^\hi)$ as in the previous section. It turns out that the most dominant component of the Riemann tensor for the metric \eqref{Bondi metric} is $R_{u\hat{A}u\hat{B}} = -\frac{1}{2r}\ddot{C}_{\hat{A}\hat{B}}+\cO(1/r^2)$, where $C_{\ha\hb}\equiv \e{A}{A}\e{B}{B}C_{AB}$. Also, $X^\hj=X_0^\hj+V_0^\hj(u-u_0)+\cO(1/r)$, and therefore  leading gravito-electromagnetic fields given in \eqref{GEM fields} obey \eqref{F asymptotic}, with
\begin{align}\label{GEM asymptotics}
	\overbar{E}_\ha=\frac{1}{2}\ddot{C}_{\hat{A}\hat{B}}\big(X_0^\hb+V_0^\hb(u-u_0)\big)\,.
\end{align}
Using this result in Eq.\eqref{memories general}, we can derive the gravitational displacement and kick memory effects. We represent the result using the notation $\deq$, which means equality up to $\cO\big((X,\dot{X})^3,r^{-2}\big)$ corrections\footnote{In deriving these results, the following identities turn out to be useful
	\begin{align}
		\cM_1({{\overbar{E}}}_\ha)&= \frac{1}{2} V_0^\hb \cM_2(\ddot{C}_{\hat{A}\hat{B}})\,,\qquad 	
		\cM_2({{\overbar{E}}}_\ha)= \frac{1}{2} \Big(X_0^\hb \cM_2(\ddot{C}_{\hat{A}\hat{B}})+ V_0^\hb \cM_3(\ddot{C}_{\hat{A}\hat{B}})\Big)\,,
	\end{align}
	and $ \cM_n(\ddot{C}_{\hat{A}\hat{B}})=-(n-1)\,\cM_{n-1}(\dot{C}_{\hat{A}\hat{B}})$ for $n\geq2$, assuming that the news vanishes when $u\geq u_f$.
}:
\begin{subequations}\label{gravitational memories}
	\begin{align}
	\Delta V^\ha&\deq -\frac{1}{2r} V_0^{\hat{B}}(1-V_0^\hr)\,\Delta C_{\hat{A} \hat{B}} \,,\label{kick memory}\\
	\Delta X^\ha&\deq \frac{1}{2r} \left[\big((1-3V_0^\hr)X_{0}^{\hat{B}}-X_0^\hr V_0^\hb\big)\Delta C_{\hat{A} \hat{B}} -2V^{\hat{B}}_0 (1-\tfrac{3}{2}V_0^\hr) \int_{u_0}^{u_f}\!\!\! du \big(  C_{\ha\hb}(u) - C_{\ha\hb}(u_f)\big)\right],\label{displacement memory} \\
	\Delta{V}^{\hr}&\deq \frac{1}{2r}\, V_0^\ha V_0^\hb\Delta C_{\hat{A} \hat{B}}\,,\\
		\Delta{X}^{\hr}&\deq\frac{3}{2r}\, V_0^\ha V_0^\hb\int_{u_0}^{u_f}\!\! du \big(  C_{\ha\hb}(u) - C_{\ha\hb}(u_f)\big)\,.
\end{align}
\end{subequations}
One can sort this result in a relativistic expansion in powers of the initial velocity $V_0^\hi$. At leading order, \ie in the  limit $|V_0|\to 0$, the only effect is is the well-known leading displacement effect \cite{Christodoulou:1991cr}. At subleading order (linear in $V_0^\hi$), there is  a `subleading' correction to the displacement memory observable, as well as a kick memory in the transverse plane \cite{Nichols:2017rqr, Nichols:2018qac, Flanagan:2018yzh,Tahura:2020vsa,Seraj:2021qja,Tahura:2021hbk}. The terminology comes from the observation that first and second terms in the displacement memory coincide with leading and subleading soft graviton currents \cite{Campiglia:2020qvc}, but  is also consistent with the result that the subleading term appears at higher order in a post-Newtonian expansion of the source \cite{Nichols:2017rqr,Nichols:2018qac}. Finally, at quadratic order in $V_0^\hi$,  radial displacement and kick memory effects show up, which explains why it has not been noted previously in the literature. Note the appearance of $\cO(X_0V_0)$ corrections in the first term of \eqref{displacement memory} and the absence of corrections of the form $\cO(X_0^2)$.

Let us compare memory effects in EM and GR, which have several similarities and differences. EM and gravitational fields both satisfy similar asymptotic behavior  represented by Eq.\eqref{F asymptotic}, and test bodies in both theories obey similar evolution equation given by \eqref{generalized Lorentz force}. At the same time, EM and GR are different in the detailed form of the transverse electric field, given by \eqref{Electric field EM} and \eqref{GEM asymptotics} respectively. Moreover, the charge to mass ratios $\beta_E,\beta_{B}$ and the subleading extra force $F_{\text{ext}}^\hi$ are different in these theories. As a result of similarities, there is a displacement and a kick effect at leading order in the $1/r$ expansion in both theories. However, as a result of differences, with respect to an expansion in the Mellin transform of the radiative field ($\dot{\overbar{A}}_\ha$ in EM and the news tensor $\dot{C}_{\ha\hb}$ in GR), the displacement effect is leading in GR, while it is subleading in EM. The kick memory is leading in both cases, however, it appears in GR only if there is a nonzero initial velocity, while in EM, it is nonzero even in the absence of initial velocity.

\subsection{Gyroscopic memory}\label{sec: gyroscopic memory}
The effect of gravitational field on gyroscopes, \ie  objects carrying spin, has been of interest for the whole history of general relativity. The famous Lens-Thirring effect explains how a rotating star/black hole affects the spin of the gyroscope, and this has been tested by the Gravity Probe B experiment \cite{Everitt:2011hp}. The effect of gravitational waves on gyroscopes has been discussed in \cite{Herrera:2000uh,Bini:2000xj,Sorge:2001sq,ValienteKroon:2001pc}, and more thoroughly recently in \cite{Seraj:2021rxd,Seraj:2022qyt} for freely falling gyroscopes. 

The evolution of a freely falling small gyroscope with velocity $V^\mu$ and spin $S^\mu$ is given by the parallel transport equation $V^\alpha\nabla_\alpha S^\mu=0$. With respect to a local frame which is comoving with the gyroscope, \ie  when $e_{\hat{0}}{}^\mu=V^\mu$, the gyroscope's spin  is purely spatial given by $S^\hi=S^\mu e^\hi{}_\mu$, while $S^{\hat{0}}=-S^\mu u_\mu=0$. The parallel transport equation then reads
\begin{align}\label{parallel transport}
	\frac{d S^{\hat{i}}}{d \tau}=\Omega^{\hat{i}}_{\;\hat{j}}  S^{\hat{j}},
	\qquad
	\Omega^{\hat{i}\hat{j}}	\equiv-V^{\alpha}\omega^{\hat i \hat j}{}_\alpha,
\end{align}
where  $\omega^{\hat \mu \hat \nu}{}_\alpha=e^\hmu{}_\beta \nabla_\alpha e^{\hnu\beta}$ is the spin connection of the local frame with respect to which the orientation of the spin is computed. Note that a free gyroscope shows no precession in a parallel transported frame. To measure a nontrivial observable effect, an optical  frame adapted to light rays arriving from distant stars was introduced in \cite{Seraj:2021rxd,Seraj:2022qyt} which reduces to \eqref{frame Minkowski} in the asymptotic limit. It was shown that the effect of GWs on a gyroscope with initial spin $S^{\ha}(u_0)$ is a  net change of orientation, given by a rotation $\Delta\Phi$ in the transverse plane orthogonal to the direction of propagation of GWs\footnote{Our convention in defining the dual shear is opposite of that of \cite{Seraj:2021rxd,Seraj:2022qyt} and matches with \cite{Godazgar:2018dvh,Freidel:2021qpz}, which causes a relative minus sign in the angle $\Delta\Phi$.}
\begin{align}\label{gyroscopic memory}
	\Delta S^{\ha}&=\eps^{\ha\hb}S_{\hb}(u_0)\Delta\Phi\,,\qquad \Delta\Phi=-\frac{1}{r^2}
	\int du \left(\tfrac14 D_{A}D_B\tl C^{AB}-\tfrac18 N_{AB}\tl C^{AB}\right) \,.
\end{align}
The first term of the integral coincides with the spin memory effect \cite{Pasterski:2015tva}, which is explicit in the formulation of spin memory in \cite{Nichols:2017rqr}. This shows the close connection between the two effects. However, there is an additional nonlinear contribution to the gyroscopic memory, related to the charge of gravitational electric-magnetic duality, discussed in \cite{Seraj:2021rxd,Seraj:2022qyt}.

\section{Memory from  holonomy}\label{sec:GR Wilson loops}
In this section, we will show that various memory effects can be derived from another nonlocal quantity, namely \textit{holonomy}, see \eg \cite{Loll:1993yz} for a nice exposition.

In the context of gauge theory, consider the parallel transport equation  over a path  $\Gamma$ parameterized by $\tau\in[\tau_0,\tau_f]$:
\begin{align}\label{parallel transport eq}
	\dfrac{D}{D\tau}W\equiv \frac{dx^\mu}{d\tau} D_\mu W=0\,, \qquad D_\mu=\pd_\mu+A_\mu\,,
\end{align}
where the connection $A_\mu=A^a_\mu \, T_a$ gauges the Lie algebra represented by generators $T_a$, and $q$ is the charge of the test particle coupled to the gauge field. For a given path $\Gamma$ (not necessarily closed), the holonomy $W(\Gamma)$, as an element of the gauge group, is the solution to \eqref{parallel transport eq} with initial condition $W[x(\tau_0)]=I$. It is given by the path ordered exponential of the connection:
\begin{align}\label{Wilson loop}
	W[\Gamma]&=\cP \exp\left( -\int_{\Gamma}dx^\mu A_\mu \right)\,,
\end{align}
where $\cP$ is the path ordering operator. Note that we use natural units where $\hbar=c=G=1$. The trace of this operator is called a Wilson line, or a Wilson loop if the path is closed. In Abelian gauge theories like Maxwell, the trace is trivial and thus the two notions coincide.

\subsection{EM memory from holonomy} 
\begin{SCfigure}
	\centering
	\caption[width=0.7\textwidth]{A closed path in the transverse plane, used to compute the EM and gravitational holonomies. The path is specified by the initial data, \ie the initial relative position and velocity $(X_0^\ha,V_0^\ha)$ and the vertical length is given by the experiment time interval $(u_0,u_f)$. Note that this figure is suppressing one direction in the transverse plane: in general $X_0^\ha,V_0^\ha$ are independent vectors. Note, however, that we assume $X_0^\hr=0=V_0^\hr$.}
	\includegraphics[width=.45\textwidth]{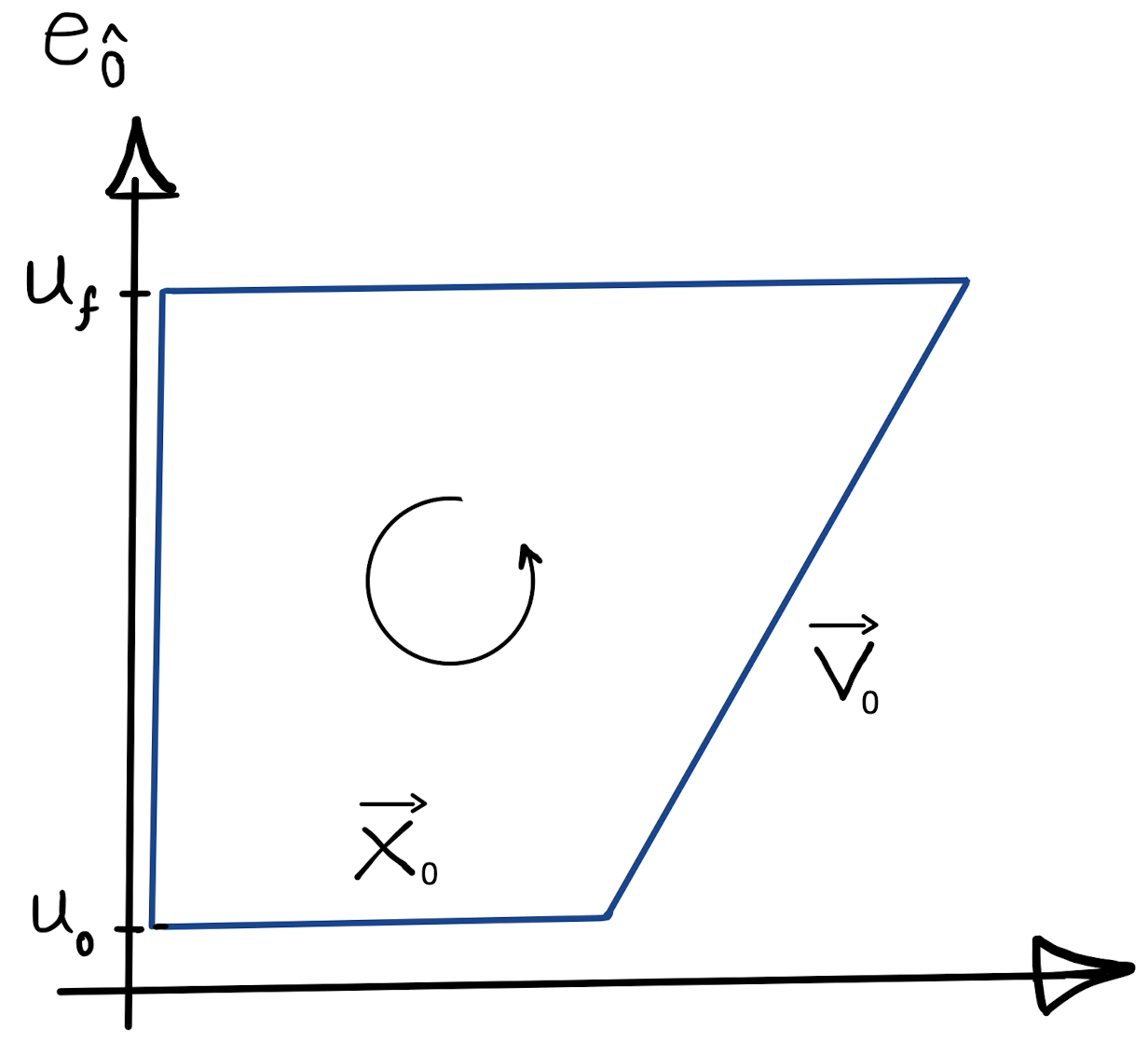}
	\label{Fig:trapezoid}
\end{SCfigure}

Let us compute the path ordered exponential \eqref{Wilson loop} along the trapezoidal spacetime loop shown in Fig.\ref{Fig:trapezoid}. The loop is specified by the initial data of the test particle: the initial position and velocity $(X^\ha_0,V^\ha_0)$ in the transverse plane orthogonal to the propagation direction. The vertical axis represents the time interval of the experiment $(u_0,u_f)$. We compute \eqref{Wilson loop} in the local orthonormal frame constructed in section \eqref{sec:Maxwell}, and we note that $T=u+\cO(1/r^2)$, so that
\begin{align}\label{holonomy EM}
	{W} &= \frac{1}{r}\left[\int_0^{X_0} dX^\ha \overbar{A}_\ha(u_0) - \int_0^{X(u_f)}  dX^\ha \overbar{A}_\ha(u_f)\right.\nonumber\\
	&\qquad\qquad\left. - \int_{u_0}^{u_f} du\, \overbar{A}_u(u) + \int_{u_0}^{u_f} du\, \left(\overbar{A}_u (u) +V_0^\ha \overbar{A}_\ha(u) \right)\right]+\cO(1/r^2)\,.
\end{align}

The first (second) line contains the contributions of the spatial (timelike) segments. Note that by assumption, the length scale of the experiment is much less than $r$, and thus in the above expression, the gauge field is essentially evaluated at a single point on the celestial sphere, while time dependence is general. As a result, the integrals over $\overbar{A}_u$ cancel out and we are left at leading order with
\begin{align}\label{holonomy displacement memory}
	{W} &= \frac{1}{r} \left[ - X^\ha_0 \Delta \overbar{A}_\ha+ V_0^\ha  \int_{u_0}^{u_f}du  \big(\overbar{A}_\ha(u) - \overbar{A}_\ha(u_f)\big)   \right]\\
	&=-\frac{1}{r} \left[ X^\ha_0 \cM_1(\dot{\overbar{A}}_\ha)+V_0^\ha \cM_2(\dot{\overbar{A}}_\ha)  \right]\,.
\end{align}
Comparing this result with \eqref{memories EM}, we observe that EM memory effects up to $\cO(|V_0|)$ can be nicely written in terms of the Wilson loop as 
\begin{align}\label{Memory-holonomy-EM}
	\Delta{X}^{\ha}=\beta\frac{\pd}{\pd V_0^\ha}{W} \,,\qquad \Delta{V}^{\ha}=-\beta\frac{\pd}{\pd X_0^\ha}{W} \,,
\end{align}
where $\beta=1-2V_0^\hr$. The dynamics of the test particle is described by the phase space described by  the pair $Z^I\equiv (X^\ha, V^\ha)$ of position and velocity. Memory effects define a tangent vector $\Delta Z^I\equiv (\Delta X^\ha,\Delta V^\ha)$, which according to \eqref{Memory-holonomy-EM} can be written as a Hamiltonian vector field
	\begin{empheq}[box=\othermathbox]{align}\label{memory-holonomy-canonical}
\Delta Z^I=\Omega^{IJ}\frac{\pd}{\pd Z_0^J}H_{\text{mem}}\,,\qquad \Omega^{IJ}=\begin{pmatrix}
			0&I\\-I&0
		\end{pmatrix}\,,\qquad H_{\text{mem}}=\beta{W}\,,
\end{empheq}
where $\Omega^{IJ}$ is the inverse symplectic form and $H_{\text{mem}}$ is the corresponding Hamiltonian.

\subsection{Gravitational holonomy: formal development}\label{sec: GR holonomy formal}
In this section, we derive the above gravitational memory effects from certain gravitational holonomies. To this end, it is convenient to think of general relativity as a gauge theory of the Poincar\'e group (see \cite{Ivanenko:1983fts} and references therein). Let us define the `Poincar\'e gauge field' as
\begin{align}\label{GR gauge field}
A_{\mu}=e^{\hmu}{}_{\mu} P_{\hmu}+\frac{1}{2} \omega^{\hmu\hnu}{}_{\mu} M_{\hmu\hnu}\,,
\end{align}
where $P_\hmu,M_{\hmu\hnu}$ respectively denote the generators of Poincar\'e translations and Lorentz transformations, respectively gauged by the vielbein $e^{\hmu}{}_{\mu}$ and the spin connection $ \omega^{\hmu\hnu}{}_{\mu}$. The explicit form of $P_\hmu,M_{\hmu\hnu}$ depends on the representation. From \eqref{GR gauge field}, we define the `Poincar\'e holonomy' as
\begin{align}\label{GR holonomy gauge field}
W[\Gamma]=\cP \exp\left( -\int_{\Gamma}dx^\mu A_\mu \right)=\cP \exp \left(W^\hmu P_{\hmu} +\frac12 W^{\hmu\hnu} M_{\hmu\hnu}\right)\,,
\end{align}
where
\begin{subequations}
	\begin{align}
W^\hmu &= -\int_{\Gamma} dx^\mu \;e^{\hmu}{}_{\mu}\;\,=\frac{1}{r}\overbar{W}^\hmu+\frac{1}{r^2}{W}_{(2)}^\hmu+\cO(1/r^3), \label{translational holonomy}\\
W^{\hmu\hnu}&=-\int_{\Gamma} dx^\mu\,\omega^{\hmu\hnu}{}_{\mu}=\frac{1}{r^2}\overbar{W}^{\hmu\hnu}+\cO(1/r^3)\,. \label{rotational holonomy}
\end{align}
\end{subequations}
The last equations in each line above come from an asymptotic expansion of fields, as we will see explicitly in section \ref{sec:source oriented frame}. Equation \eqref{GR holonomy gauge field} can be simplified using the Zassenhaus formula, stating that for two operators $X,Y$
\begin{align}
e^{X+Y}=e^{ X} e^{ Y} e^{-\frac{1}{2}[X, Y]} e^{\frac{1}{6}(2[Y,[X, Y]]+[X,[X, Y]])} \cdots
\end{align}
where $\cdots$ refer to exponentials involving more commutators. Taking $X=W^\hmu P_{\hmu}, Y=\frac12 W^{\hmu\hnu} M_{\hmu\hnu}$, and  given the asymptotic form of $W^\hmu,W^{\hmu\hnu}$, we find that
\begin{align}
	W&=\cP\left(\exp (W^\hmu P_{\hmu}) \exp( \frac12 W^{\hmu\hnu} M_{\hmu\hnu})\right)+\cO(1/r^3)\\
	&=1+{W}^\hmu P_\hmu+\frac{1}{2}{W}^{\hmu\hnu}  M_{\hmu\hnu}+\frac{1}{2}\cP({W}^\hmu{W}^\hnu) P_\hmu P_\hnu+\cO(1/r^3)\,. \label{holonomy leading}
\end{align}
\paragraph{Finite dimensional representation.} We can explicitly compute the holonomy once we have a specific representation of the Poincar\'e algebra. In a $d$ dimensional spacetime, a finite  representation is given by a $d+1$ dimensional vector $\Psi^a$, which transforms under the Poincar\'e group as
\begin{align}\label{Poincare group rep}
	\Psi^a\equiv \begin{pmatrix}
		\psi^\hmu \\
		c 
	\end{pmatrix}\mapsto \begin{pmatrix}
	\L^\hmu{}_\hnu & T^\hmu \\
	0&  1 
\end{pmatrix}
	\begin{pmatrix}
		\psi^\hmu \\
		c 
	\end{pmatrix} \,,
\end{align}
where $c\in \mathbb{R}$ is an arbitrary real number. Note that $c$ is invariant under the symmetry transformation and therefore the representation is decomposed into various orbits labeled by $c$, while $\psi^\hmu$ represents the physical state of the system. Depending on the value of $c$, $\psi^\hmu$ represent different physical systems. In particular, on the orbits $c=1,c=0$, the state transforms as  
\begin{align}
X^\hmu&\equiv\psi^\hmu[c=1]\,,	&&\hspace{-2.1cm}X^\hmu\mapsto  \L^\hmu{}_\hnu \,X^\hnu+T^\hmu\,,\\
S^\hmu&\equiv\psi^\hmu[c=0]\,,	&&\hspace{-2cm}S^\hmu\mapsto  \L^\hmu{}_\hnu \,\psi^\hnu\,.
\end{align}
$X^\hmu$ can represent the position of a particle in the local frame, while $S^\hmu$ can represent the spin of a gyroscope.  By linearizing \eqref{Poincare group rep} around the identity, we can find the matrix representation of the Poincar\'e algebra
\begin{align}\label{Poincare algebra rep}
	(M_{\hmu\hnu})^a{}_b=\de_{\hmu}{}^a \eta_{\hnu b}-\de_{\hnu}{}^a \eta_{\hmu b}\,\qquad (P_{\hmu})^a{}_b =\de_{\hmu}{}^a\,\de_b{}^{d+1}\,.
\end{align}
In particular, translation $P_\hmu$ is nilpotent, \ie $P^n=0$ for $n\geq2$ and therefore, the holonomy \eqref{holonomy leading} reduces to 
\begin{align}\label{GR holonomy simplified}
	W&=1+\frac{1}{r}\,\overbar{W}^\hmu P_\hmu+\frac{1}{r^2}\left(\frac12\overbar{W}^{\hmu\hnu}  M_{\hmu\hnu}+W_{(2)}^\hmu P_\hmu\right)+\cO(1/r^3)\,.
\end{align}
Therefore, at leading order in $1/r$ expansion, the holonomy is purely translational given by $\overbar{W}^\hmu$, while it consists of a rotational (Lorentz) part $\overbar{W}^{\hmu\hnu}$ as well as a subleading translational part $W_{(2)}^\hmu$ at $\cO(1/r^2)$. In fact, it turns out that the only nontrivial components at leading order are $\overbar{W}^\ha, \overbar{W}^{\ha\hb}$, \ie  when they involve transverse directions. We show that $\overbar{W}^\ha$ encodes the (leading and subleading) displacement memory, while the rotational holonomy $\overbar{W}^{\ha\hb}$ reproduces the gyroscopic memory once integrated along appropriate paths. 

\paragraph{Poincar\'e holonomy and affine transport equation.} A natural question is: which transport equation does the Poincar\'e holonomy solve? To answer this question, we consider again the finite dimensional representation discussed above. Assume that the initial state of the probe system is given by $\Psi^a_0$. Now the evolution of the initial state by the holonomy \eqref{GR holonomy gauge field} is given by $\Psi^a(\tau)=(W_\tau)^a{}_b \Psi_0^b$, where $W_\tau$ is the holonomy along an open path with endpoint $x^\mu(\tau)$ and with tangent vector $k^\mu=dx^\mu/d\tau$. It satisfies the differential equation
\begin{align}
	\dfrac{d}{d\tau}\Psi^a=-k^\mu (A_\mu)^a{}_b \Psi^b\,.
\end{align}
Using the decomposition $\Psi^a=(\psi^\hmu,c)$, and equations \eqref{GR gauge field},\eqref{Poincare algebra rep}, we find that the evolution equation, depending on the value of $c$, is given by 
\begin{align}
\dfrac{d}{d\tau}\psi^\hmu=-c k^\hmu-k^\mu\omega_\mu{}^\hmu{}_\hnu \psi^\hnu \,.
\end{align}
Taking the last term to the left hand side, and multiplying the equation with inverse tetrad to go back to the coordinate basis reveals
\begin{align}
k^\mu \cd_{\mu} \psi^\nu=-c \, k^\nu\,\qquad \psi^\mu\equiv e_\hmu{}^\mu\,\psi^\hmu\,.
\end{align}
Therefore, on the orbit $c=0$, the evolution reduces to the parallel transport equation, while for $c=1$, we recover the `affine' transport equation of Refs. \cite{Flanagan:2014kfa,Flanagan:2018yzh}. The latter case ($c=1$) corresponds to the evolution of the position of a test mass in the local frame, or equivalently, the physical distance between two nearby test masses, while the former case ($c=0$) describes the evolution of the spin of a gyroscope.

\subsection{Asymptotic frame} \label{sec:source oriented frame}
To compute the holonomy \eqref{GR holonomy simplified}, we need to pick a particular frame. In this paper, we use the `source oriented frame' constructed in \cite{Seraj:2021rxd,Seraj:2022qyt,Godazgar:2022pbx}, which is orthonormal everywhere in spacetime and therefore provides an extension of the asymptotic frame \eqref{frame Minkowski} inside the bulk. It is constructed such that the timelike basis vector $\bolde_{\hat 0}$ coincides with a geodesic congruence that is at rest at some initial time $u_0$. Moreover, the radial basis vector $\bolde_{\hat r}$ is tangent to the spatial path followed by rays arriving from the source. However, as we will discuss later, the essence of our result, namely the fact that holonomies encode memory effects is independent of the choice of frame. To support this, we study the gauge transformation properties of the translational holonomy $\overbar{W}^\ha$. 

At large distance, the dual basis one-forms of the source oriented frame were derived in \cite{Seraj:2021rxd,Seraj:2022qyt} which in the large distance limit take the form
\begin{align}\label{source oriented frame}
	\bolde^{\hat{0}}&=\dd u+\dd r\,,\quad
	\bolde^{\hat{r}}=\dd r\,,\quad
	\bolde^{\hat{A}}=E^{\ha}{}_{B}\Big(\frac{1}{r}D_CC^{BC} \dd u+(r\de^B_{\;\;A }+\frac{1}{2}C^B_{\;\;A})\,\dd\th^A\Big)
\end{align}
where $E^{\ha}{}_{A}(\th^A)$ is a time-independent dyad on the sphere as in section \ref{sec:Maxwell}. The associated spin connection is given by $\widetilde\omega=\overbar{\omega}+\omega$ where 
\begin{align}
	\overbar{\omega}_{\hr\ha}
	=
	-E_{\ha A}d\th^A,
	\qquad
	\overbar{\omega}_{\ha\hb}
	=
	\barE{A}{B}D_B E_{\hb C}d\th^C,
\end{align}
and up to irrelevant subleading corrections
\begin{align}
	\label{spin connection}
	\omega_{\hr\ha}
	&=
	\barE{A}{A}\left(\frac{1}{4r^2} N_{AB}D_C C^{BC}du
	+\frac{1}{2r^2}D^BC_{AB}dr
	+\frac{1}{2}N_{AB}d\th^B\right),\\
	\label{e34}
	\omega_{\ha\hb}
	&=\eps_{\ha\hb}\left[
	-\frac{1}{4r^2}\big(D_AD_B \tl C^{AB}-\frac12 N_{AB}\tl C^{AB}\big)du+\frac{1}{2r}D^A\tl C_{AB} d\th^B\right]\,.
\end{align}
In the above result, the $\overbar{\omega}$ part of the spin connection is nonzero even in Minkowski background, while the genuine radiation effects are encoded in $\omega$. Since the spin connection does not transform covariantly, it is possible to find a rotated frame in which the connection is solely given by $\omega$. It was shown in \cite{Seraj:2021rxd,Seraj:2022qyt} that the latter corresponds to a frame whose spatial axes are tied to distant stars rather than being source-oriented. 

In the following subsections, we first compute the translational holonomy over the path depicted in Fig.\ref{Fig:trapezoid} and show that it reveals the displacement and kick memory effects. Then, we compute the rotational holonomy over the paths depicted in Fig.\ref{fig:GR loops} and show that they reproduce the gyroscopic and spin memory effects discussed in section \ref{sec:GR memory effects}. 

\subsection{Translational holonomy}\label{sec: translational holonomy}
Consider two test masses with initial relative physical distance and velocity $X^\ha_0\equiv X^\ha(u_0)$ and $ V^\ha_0\equiv V^\ha(u_0)$ in the transverse plane. We can think of this as the position and velocity of the second test mass in the local inertial frame \eqref{source oriented frame} in which the first mass defines the origin. Also, assume that the GW is being emitted during the time interval $(u_0,u_f)$. From this data, we construct a trapezoid depicted in figure \ref{Fig:trapezoid}, and compute the translational holonomy \eqref{translational holonomy}. The computation is the same as in \eqref{holonomy EM}, except that we replace $A_\mu$ by $e^\ha{}_\mu$, given by   \eqref{source oriented frame}. The result at leading order is 

\begin{empheq}[box=\othermathbox]{align}\label{holonomy displacement memory}
	W_\ha = -\frac{1}{2r} \left[  X^{\hat{B}}_0 \Delta C_{\ha\hb} - V_0^{\hb}  \int_{u_0}^{u_f}du  \big(C_{\ha\hb}(u) - C_{\ha\hb}(u_f)\big)   \right]+\cO(1/r^2)\,.
\end{empheq}
Restricting attention to memory effects at most linear in $V_0^\hi$ in \eqref{gravitational memories}, we can represent nontrivial displacement and kick memory observables in terms of the translational holonomy as 
\begin{align}\label{Memory-holonomy-GR}
	\Delta{X}_{\ha}=-\big[\beta X_0^\hb\frac{\pd}{\pd X_0^\hb}+\beta_{B}V_0^\hb\frac{\pd}{\pd V_0^\hb}\big]{W}_\ha \,,\qquad \Delta{V}_{\ha}=\beta_{E}V_0^\hb\frac{\pd}{\pd X_0^\hb}{W} _\ha\,,
\end{align}
where $\beta=1-3V_0^\hr$. This provides an explicit relationship between the leading part of the translational holonomy and gravitational displacement and kick memory observables. We have not been able to identify a canonical structure in the above result, similar to Eq.\eqref{memory-holonomy-canonical} for EM. We leave a detailed analysis of this issue to a future work.

\subsection*{Remarks.}

\emph{Composition of holonomies.} For two paths $\Gamma_1,\Gamma_2$, the composition $\Gamma_{1} \circ \Gamma_{2}$ is another loop in which overlapping segments with reverse arrows cancel out. The composition rule states that 
\begin{align}
	W_{\Gamma_{1} \circ \Gamma_{2}}(A)=W_{\Gamma_{2}}(A) \cdot W_{\Gamma_{1}}(A), \quad \forall A,
\end{align}
The trapezoid in figure \eqref{Fig:trapezoid} can be decomposed into a rectangular loop and a triangular loop. The former encodes the first term in \eqref{holonomy displacement memory}, \ie  the leading displacement memory, while the triangular loop encodes the subleading displacement memory.

\noindent\emph{Gauge transformation of the holonomy.} 
We computed the holonomy above in a specific frame given by \eqref{source oriented frame}. However, the choice of frame is arbitrary and parameterized by internal Lorentz symmetries. In contrast to the Wilson loop, holonomy is not gauge invariant, but transforms under a gauge transformation $A\to g\cdot A=g^{-1}A g+g^{-1}\dd g$ as
\begin{align}
	W[\Gamma,g\cdot A]=g^{-1}[\tau_f]\,W[\Gamma,A]\,g[\tau_0]\,,
\end{align}
where $\tau_0,\tau_f$ refer to the starting and end points of the path $\Gamma$. Now let us see how our result above transforms at leading order in $1/r$ under an infinitesimal Local Lorentz transformation $g=I+T^\ha P_\ha+\lambda^{\ha\hb}M_{\ha\hb}$. At leading order, $W[\Gamma, A]=1+ \frac{1}{r}\overline{W}^\ha P_\ha +\cO(1/r^2)$, and  therefore the gauge transformation induces
\begin{align}
	\overbar{W}^\ha P_\ha\to \overbar{W}^\ha\left( (\delta_\ha{}^\hb+\l_\ha{}^\hb)P_\hb+\l^{\hat{0}}{}_\ha P_{\hat{0}}\right)\,,
\end{align}
implying that the translational holonomy $W_{(1)}^\ha$ is a Lorentz vector, rotated by $\l^{\hat A\hat B}$ and boosted by $\l^{\hat{0}\ha}$. This is expected since `displacement' memory also transforms as a vector.

\noindent\emph{Time translation.} Intuitively, we may expect that  $W^{\hat{0}}$, \ie  the time-translational holonomy gives the so called ``relative proper time memory" \cite{Strominger:2014pwa,Flanagan:2018yzh}. However, computing this quantity, we find that 
\begin{align}\label{relative time memory}
	W^{\hat{0}}&=\cO(1/r^3)\,.
\end{align}
This result matches that of Strominger and Zhiboedov \cite{Strominger:2014pwa} for freely falling observers. Computing the first nontrivial term in the expansion of  $W^{\hat{0}}$, requires a more accurate expansion of the frame \eqref{source oriented frame}, which goes beyond the scope of this paper. However, it would be nice to compute this and compare it with Eq.(2.6) of \cite{Flanagan:2018yzh}.

\subsection{Rotational holonomy }\label{sec:rotational holonomy}
\begin{figure}
	\captionsetup{width=.9\linewidth}
	\centering
	\begin{subfigure}[b]{0.28\textwidth}
		\centering
		\includegraphics[width=\textwidth]{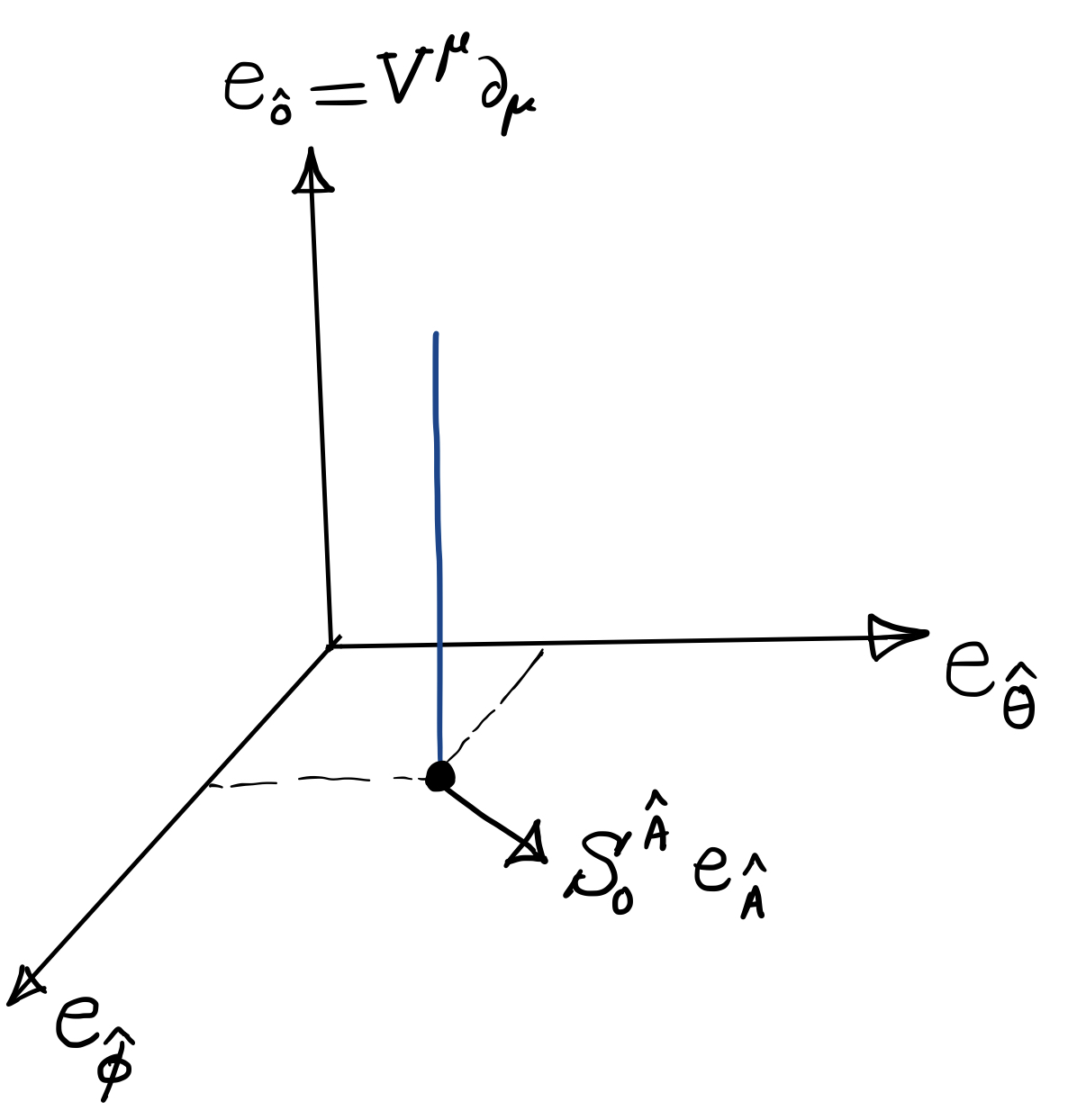}
		\caption{}
		\label{fig:gyro}
	\end{subfigure}
	\qquad
	\begin{subfigure}[b]{0.28\textwidth}
		\centering
		\includegraphics[width=\textwidth]{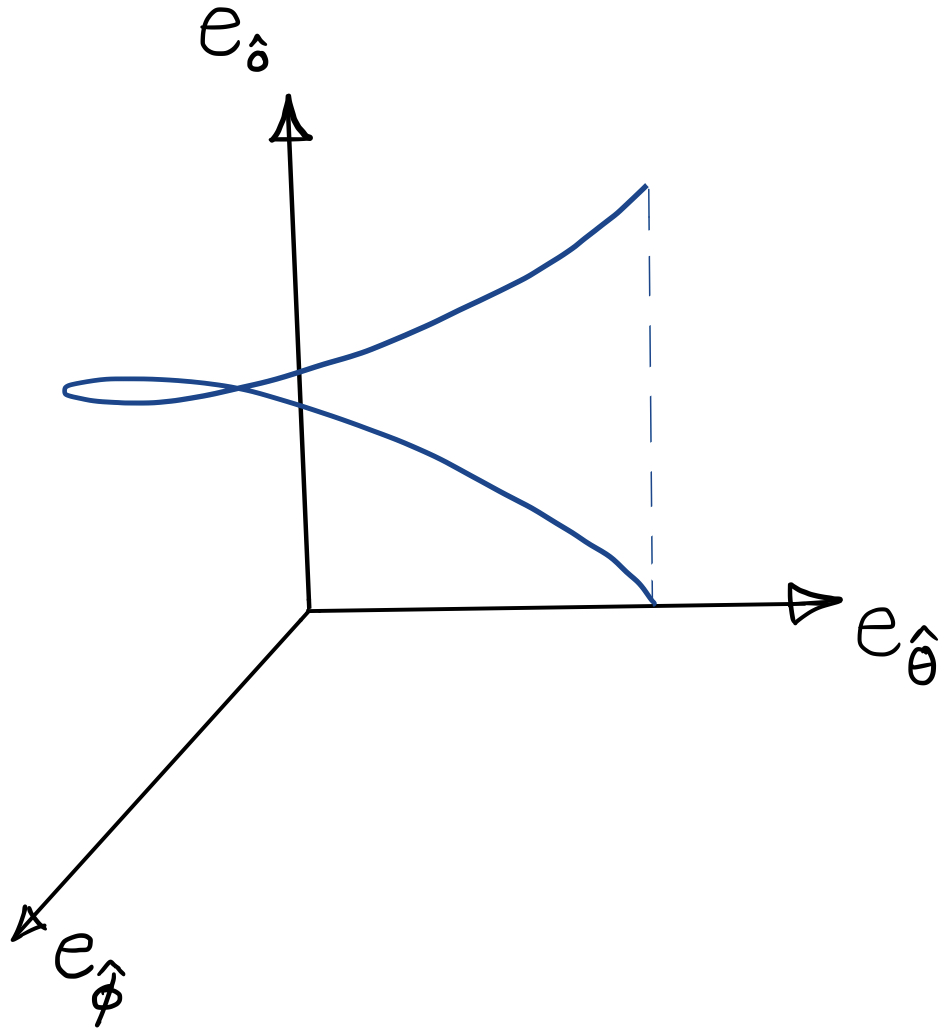}
		\caption{}
		\label{fig:helical}
	\end{subfigure}
	\hfill	
	\caption{Paths implemented to compute rotational holonomies. The paths are non-dynamical and specified by the initial data, while the dynamics is encoded in the holonomy. Figure \eqref{fig:gyro} is a timelike worldline, which appears as a straight line in the comoving frame \eqref{source oriented frame}.  Figure \eqref{fig:helical} is a helical path whose projection on transverse directions forms a loop. Both paths reproduce the gyroscopic memory.}
	\label{fig:GR loops}
\end{figure}

In this section, we study the rotational holonomy \eqref{rotational holonomy} and show that for suitable choice of the path, it reproduces the gyroscopic memory effect. 
Consider a generic path at large distance in the transverse plane. Using \eqref{spin connection}, we find that the effect of radiation on the rotational holonomy is given by
\begin{align}\label{W2ab}
{W}^{\ha\hb}=-\frac{1}{2r^2}\eps^{\ha\hb}\left[\int du \big( D_A D_B \tl{C}^{AB}-\frac{1}{2}N_{AB} \tl{C}^{AB}\big)-2\int r d\th^A D^B \tl{C}_{AB} \right]+\cO(1/r^3)\,.
\end{align}
If the path is a timelike geodesic, as in figure \eqref{fig:gyro}, which has zero velocity at some initial time, the angular velocity remains at $\cO(1/r^2)$ at later times \cite{Seraj:2021rxd,Seraj:2022qyt}. Therefore, the last term in brackets in \eqref{W2ab} is subleading and we find at leading order
\begin{align}\label{W2ab simplified}
\overbar{W}^{\ha\hb}=-\frac{1}{2r^2}\eps^{\ha\hb}\left[\int du\big( D_A D_B \tl{C}^{AB}-\frac{1}{2}N_{AB} \tl{C}^{AB}\big) \right]\,,
\end{align}
which coincides with the gyroscopic memory effect \eqref{gyroscopic memory}. This result is expected, because the evolution of a freely falling gyroscope is given by the parallel transport equation \eqref{parallel transport}, which is actually solved by the holonomy 
\begin{align}
	S^\hi=\cP \exp\left( -\int_{\Gamma} dx^\mu \omega_\mu^{\;\;\hat i}{}_\hj\right) S_0^{\hj}\,,
\end{align}
over a geodesic $\Gamma$ followed by the gyroscope. The net change in the spin is given by $S^{\hj}(u_f)-S_0^{\hj}$ which is at leading order given by minus the rotational holonomy $W^{\hi\hj}$ defined in \eqref{rotational holonomy}. In particular, at $O(1/r^2)$, the only effect appears in $\hi\hj=\ha\hb$ components, which gives the gyroscopic memory, \ie  a rotation in the transverse plane, given by \eqref{gyroscopic memory}.

Another interesting situation is when the path is not a geodesic, but instead its spatial projection forms a loop in the transverse plane, as depicted in figure \eqref{fig:helical}. In this case, the angular integral in \eqref{W2ab} can also be written as a surface integral 
\begin{align}\label{W2abclosed}
	\overbar{W}^{\ha\hb}=-\frac{1}{2r^2}\eps^{\ha\hb}\left[\int du\big( D_A D_B \tl{C}^{AB}-\frac{1}{2}N_{AB} \tl{C}^{AB}\big) -r\int_{D(\Gamma)} d^2\th \sqrt{q} D_AD_B {C}^{AB} \right]\,.
\end{align}
Since the size of the path, denoted by $L$, is by assumption much smaller than the distance $r$ to the source, the angular integral is subleading by a factor $L/r$ with respect to the first integral (since $d\theta\sim L/r$),  and we again reproduce the gyroscopic memory \eqref{W2ab simplified}.

\section{Discussion and outlook}\label{sec:discussion}
In this paper, we showed that known memory effects can be derived from certain Wilson loops/ holonomies on a holographic screen, \ie  a hypersurface of constant radius at large distance in asymptotically flat spacetimes. There are various ways to extend this result:
\begin{itemize}[leftmargin=*]
	\item One may extend this construction to asymptotically AdS spacetimes, where holonomies play a prominent role in AdS/CFT. In this setup, Wilson loops in the field theory side correspond to the area of certain minimal surfaces in the bulk, and a phase transition occurs when the minimal surface starts to meet the black hole horizon \cite{Ammon:2015wua}. 
	\item Using the Nonabelian Stokes theorem \cite{schreiber2011smooth} which generalizes the usual Stokes theorem to Lie algebra valued differential one-forms, one may be able to write equivalent expressions for the holonomies that we derived as dual surface integrals. This will be analogous to what happens in AdS/CFT. 
	\item From the boundary point of view, it is believed that gravity in asymptotically flat spacetimes is holographically described by a Carrollian field theory \cite{levy1965nouvelle,Duval:2009vt,Duval:2014uva,Ciambelli:2018wre,Bagchi:2019xfx,Herfray:2021qmp,Donnay:2022aba}. If such a correspondence exists, holonomies discussed here may be related to `out of time ordered correlators' (OTOC) on the boundary. Such observables have already received much attention in the context of AdS/CFT \cite{Maldacena:2015waa}. Further investigations along these lines will be of great interest.
	\item The gyroscopic memory discussed here, based on  \cite{Seraj:2021rxd,Seraj:2022qyt}, is the net rotation, caused by GWs, of a gyroscope with respect to an optical frame constructed using light rays from distant stars. However, the fact that it can be represented as a holonomy which is closely related to the homogeneous part of the holonomy discussed in \cite{Flanagan:2014kfa,Flanagan:2018yzh}, suggests that the gyroscopic memory might be closely related to the rotation memory introduced in those references as the net change in the orientation of two nearby gyroscopes.
	\item Finally, it would be interesting to study more subleading memory effects and their relationship to holonomies, in particular, the relative proper time memory mentioned around Eq.\eqref{relative time memory}. 
	
\end{itemize}

\section*{Acknowledgments}
AS is grateful to Aldo Riello for proposing the possible relationship between Wilson loops and memory effects. We are grateful to the anonymous referee who helped us greatly improve the paper, and in particular section \ref{sec:GR memories}. We also thank H. Afshar, W. Merbis, B. Oblak, Jakob Salzer and M.M. Sheikh Jabbari for useful discussions. This research is supported by the
ERC Advanced Grant “High-Spin-Grav”, and the ERC Starting Grant ``Holography for realistic black holes''.

\addcontentsline{toc}{section}{References}

\providecommand{\href}[2]{#2}\begingroup\raggedright\endgroup

\end{document}